\documentclass[twocolumn,prc,aps,showpacs,superscriptaddress,amsmath,amssymb,floatfix,nofootinbib,superscriptaddress]{revtex4}

\usepackage{amsmath}
\usepackage{bm}
\usepackage{dcolumn}
\usepackage{subfigure}
\usepackage{amssymb}
\usepackage{xspace}
\usepackage{graphicx}
\usepackage{mathrsfs}
\usepackage{epsfig}
\usepackage{multirow}

\usepackage{amssymb}
\usepackage{amsmath}
\usepackage{graphicx} % Include figure files
\usepackage{epsfig}   % Include figure files
\usepackage{dcolumn}  % Align table columns on decimal point
\usepackage{rotating} 
\usepackage[dvips]{color}

\usepackage{float}    % To make bordet aroung figures
\floatstyle{boxed}
\usepackage{cases}
\usepackage{hyperref}
\usepackage{breakurl}

\begin{document}

\newcommand{\bd}[1]{\mathbf{#1}}
\newcommand{\Eq}[1]{Eq.~(\ref{#1})}
\newcommand{\Eqn}[1]{Eq.~(\ref{#1})}
\newcommand{\Eqns}[1]{Eqns.~(\ref{#1})}
\newcommand{\Figref}[1]{Fig.~\ref{#1}}

%\vspace{1.2in} NT@UW-14-05

\title{Kaon transverse charge density from space- and timelike data}

\newcommand*{\CUA}{The Catholic University of America, Washington, DC 20064, USA}
\newcommand*{\JLAB}{Thomas Jefferson National Accelerator Facility, Newport News, VA 23606, USA}
\newcommand*{\UW}{University of Washington, Seattle, WA 98195, USA}

\author{N.A.~Mecholsky} 
\affiliation{\CUA}

\author{J.~Meija-Ott} 
\affiliation{\CUA}

\author{M.~Carmignotto} 
\affiliation{\CUA}

\author{T.~Horn} 
\affiliation{\CUA}
\affiliation{\JLAB}

\author{G.A.~Miller}
\affiliation{\UW}

\author{I.L.~Pegg} 
\affiliation{\CUA}

\newpage
\date{\today}
\newpage

%%%%%%%%%%%%%%%%%%
%%%% Abstract %%%%
%%%%%%%%%%%%%%%%%%

\begin{abstract}
We used the world data on the kaon form factor to extract the transverse kaon charge density using a dispersion integral of the imaginary part of the kaon form factor in the timelike region. Our analysis includes recent data from $e^+e^-$ annihiliation measurements extending the kinematic reach of the data into the region of high momentum transfers conjugate to the region of short transverse distances. To calculate the transverse density we created a superset of both timelike and spacelike data and developed an empirical parameterization of the kaon form factor. The spacelike set includes two new data points we extracted from existing cross section data. We estimate the uncertainty on the resulting transverse density to be 5\% at $b$=0.025 fm and significantly better at large distances. New kaon data planned with the 12 GeV Jefferson Lab may have a significant impact on the charge density at distances of $b<$ 0.1fm. 

\end{abstract}

\keywords{elastic form factors, $K$-meson, non-perturbative QCD, parton distribution amplitudes}
\maketitle

%%%%%%%%%%%%%%%%%%
%%%% SECTIONS %%%%
%%%%%%%%%%%%%%%%%%

\section{Introduction}
\label{sec:intro}

Pions and kaons occupy a special role in nature~\cite{Horn:2016rip}. The pion is the lightest quark system, with a single valence quark and a single valence antiquark. It is also the particle responsible for the long range character of the strong interaction that binds the atomic nucleus together. A general belief is that the rules governing the strong interaction are left-right, i.e., chirally, symmetric.  If this were true, the pion would have no mass. The chiral symmetry of massless Quantum Chromodynamics (QCD) is broken dynamically by quark-gluon interactions and explicitly by inclusion of light quark masses, giving the pion and kaon mass. The pion and kaon are thus seen as key to confirm the mechanism that dynamically generates nearly all of the mass of hadrons and central to the effort to understand hadron structure.

The importance of the pion and kaon is evident in experimental and theoretical efforts, e.g., in measurements of their form factors~\cite{Frazer:1959zz,Farrar:1979aw,Efremov:1979qk,Nesterenko:1982gc,Amendolia:1986wj,Amendolia:1984nz,Bebek:1974ww,Bebek:1976qm,Bebek:1977pe,Ackermann:1977rp,Brauel:1979zk,Volmer:2000ek,Tadevosyan:2007yd,Horn:2006tm,Horn:2007ug,Blok:2008jy,Huber:2008id,Beilin:1987jm,Amendolia:1986ui,Dally:1980dj,Blatnik:1978wj,Mohring:2002tr,Coman:2009jk,Horn:2012zza,Chang:2013nia,Horn:2016rip}. The last decade saw a dramatic improvement in precision of charged pion form factor data and new results have become available on the transition form factor. L/T separated cross-section data that allow for the extraction of the kaon's elastic form factor, $F_K(Q^2)$ have been obtained at spacelike momentum transfers up to about $Q^2$=2.35~GeV$^2$~\cite{Mohring:2002tr,Coman:2009jk}, new measurements are planned with the 12 GeV Jefferson Lab~\cite{E12-07-105,E12-09-011}, and extensions are envisioned with a future Electron-Ion Collider (EIC). 

The concept of transverse charge densities~\cite{Soper:1976jc,Miller:2010nz} allows one to relate hadron form factors to their fundamental quark/gluon structure in QCD. They describe the distribution of charge and magnetization in the plane transverse to the direction of motion of a fast hadron. They are related to the partonic picture provided by the Generalized Parton Distributions (GPDs)~\cite{Diehl:2003ny,Radyushkin:1996nd,Radyushkin:1996ru,Goeke:2001tz,Belitsky:2005qn} that encode correlations between longitudinal momentum and transverse position, key properties of the nucleon. In general, GPDs can be understood as spatial densities at different values of the longitudinal momentum of the quark. Proton and pion transverse charge densities have been extracted from timelike~\cite{Miller:2010tz,Miller:2011du} and spacelike~\cite{Venkat:2010by,Carmignotto:2014rqa,Miller:2009qu} data. In the latter the extension to spacelike domain is accomplished by the use of dispersion relations and models to obtain separate real and imaginary parts. 

The goal of the present paper is to evaluate the world's kaon form factor data to extract the corresponding transverse charge density. Examining the current timelike data requires forming a superset with a single global uncertainty, taking into account the individual uncertainties and any differences in the form factor extraction method. This is done in section~\ref{sec:fk_parm}. We use the method of Ref.~\cite{Bruch:2004py} to parametrize the form factor data, but also include the new data from Refs.~\cite{Pedlar:2005sj,Seth:2012nn}. In the spacelike region we extracted new kaon form factor values from the L/T separated cross section data of Ref.~\cite{Coman:2009jk} using the technique successfully applied in pion form factor extractions~\cite{Horn:2016rip,Horn:2006tm,Huber:2008id,Horn:2007ug}. The extraction of the transverse density using a dispersion integral and the imaginary part of the form factor is described in section~\ref{sec:rho_extr}. Our procedure follows that used for the pion in Ref.~\cite{Miller:2011du}. Results for the kaon transverse density are presented in section~\ref{sec:transverse-density} and compared to those of the proton, pion and neutron in section~\ref{sec:transverse-density-compare}. The impact of future experiments is assessed in section~\ref{sec:transverse-density-future}.

\section{parameterization of the kaon form factor}
\label{sec:fk_parm}

The kaon's elastic electromagnetic structure is parameterized by two (charged and neutral) form factors, $F_{K}$, which depend on $t=Q^2=-q^2$, where $q^2$ is the four-momentum squared of the virtual photon. $F_{K}$ is well determined up to spacelike momentum transfers of $Q^2$ of 0.10 GeV$^2$ by elastic $K-e$ scattering~\cite{Dally:1980dj,Amendolia:1986ui}, from which the mean charge radius, $<r_K^2>$=0.34 $\pm$ 0.05 fm$^2$, has been extracted. At higher spacelike momenta the kaon form factor can, in principle, be extracted from kaon electroproduction data. A review on the extraction of meson form factors from electroproduction data can be found in Ref.~\cite{Horn:2016rip}. However, to date there are no published extractions of the spacelike kaon form factor from electroproduction data. In the timelike regime, the kaon form factor has been measured by annihilation $e^+ e^- \rightarrow K^+ K^-$ up to values of $Q^2$=17.4 GeV$^2$ (center of mass energy, $\sqrt{s}$=4.2 GeV) ~\cite{Achasov:2000am,Akhmetshin:1995vz,Ivanov:1981wf,Dolinsky:1991vq,Bisello:1988ez,Pedlar:2005sj,Seth:2012nn}. Our analysis thus primarily focuses on the evaluation of timelike kaon form factor data although spacelike data are included in the analysis.

Kaon form factor data in the timelike region have been obtained with the CMD-2 detector at the $e^+-e^-$ collider at the Budker Institute of Nuclear Physics in Novosibirsk. Measurements of the cross section of the annihiliation $\phi \rightarrow K^+ K^-$ allowed for extracting the timelike kaon form factor up to center of mass energies $\sim$ 2.1 GeV. More recent data are available from the CLEO experiment up to center of mass energies of $\sim$ 4.2 GeV. 

To describe all available timelike kaon form factor data, we developed a parameterization based on that of Ref.~\cite{Bruch:2004py}, which describes the high-energy region by a pattern of resonances consistent with QCD asymptotic behavior. In this parameterization, the timelike kaon form factor is assumed saturated by the $\rho$, $\omega$, and $\phi$ and their radial excitations,
\begin{equation}
\label{eqn-bruch-fk}
F_K(s)=\sum_{\substack{V=\rho,\omega,\phi,\\ \rho^\prime,\omega^\prime,\phi^\prime, ...}} 
\frac{\kappa_V f_V g_{V K \bar{K}}m_V}{m^2_V - s - i m_V \Gamma_V},
\end{equation}
where $\kappa_V$ is a coefficient reflecting the valence quark content of the mesons with ideal mixing, $f_V$ is the meson decay constant, $g_{V K \bar{K}}$ denotes the strong coupling contributions to the kaon form factor with various flavour contributions, and $m_V$ and $\Gamma_V$ are the meson mass and width respectively. The strong $VK \bar{K}$ coupling diagrams are distinguished by the presence and position of the $s$ quarks. Assuming isospin symmetry, there are three types of terms with: 1) no strange quarks, 2) both s and $\bar{s}$ in the $K\bar{K}$ state only, and 3) both $s$ and $\bar{s}$ in the $V$ and the $K\bar{K}$ state. Using these, the kaon form factor can be expressed in terms of vector meson contributions as~\cite{Bruch:2004py}
\begin{equation}
\label{eqn-bruch-fk-decomposed}
\begin{split}
F_K(s) & =   
\frac{1}{2} (c^K_\rho BW_\rho(s) + c^K_{\rho^\prime} BW_{\rho^\prime}(s)+c^K_{\rho^{\prime \prime}} BW_{\rho^{\prime \prime}}(s)) \\
 & + 
\frac{1}{6} (c^K_\omega BW_\omega(s) + c^K_{\omega^\prime} BW_{\omega^\prime}(s)) \\
  & + 
\frac{1}{3} (c^K_\phi BW_\phi(s) + c^K_{\phi^\prime} BW_{\phi^\prime}(s) +c^K_{\phi^{\prime \prime}} BW_{\phi^{\prime \prime}}(s))
\end{split}
\end{equation}
Here, $c^K_V$ are normalization constants denoting products of the meson decay constants and strong couplings, which have to be fitted together, and $BW$ denotes the Breit-Wigner type parameterization formulas, defined below. The widths of the $\rho$ and $\phi$ have an energy dependence. The width of the $\omega$ is assumed to be constant. Note that due to the limiting value of the BW functions at $s=0$, $F_K(0) = 1$, implying charge normalization, a model independent constraint. Following Ref.~\cite{Bruch:2004py}, we express the energy dependence of the widths as
\begin{equation}
\label{eqn-bruch-widths}
\Gamma_V (s) = \frac{m_V^2}{s} \left(\frac{p^K(s)}{p^K(m_V^2)}\right)^3 \Gamma_V
\end{equation}
The requirement that the spectral function, $\Theta(s-4m^2_t)$, vanishes below threshold translates into a constraint on the widths. In particular, the kaon momentum, $p^K(s)=(s-4m^2_K)^{1/2}/2$, must be real. To fulfill this criterion, we use a $\Theta$ function in $s$ to set $p>m_K^2$.

Three different Breit-Wigner functions were used in Ref.~\cite{Bruch:2004py} and are also employed in the present work: 
\begin{align}\label{BWfixed}
BW(s) &= \frac{m^2}{m^2 - s - i \sqrt{s} \Gamma},\\ \label{BWKS}
BW^{\textrm{KS}}(s) &= \frac{m^2}{m^2 - s - i \sqrt{s} \Gamma(s)},\\ \label{BWGS}
BW^{\textrm{GS}}_{\textrm{K or $\pi$}}(s) &= \frac{m^2+H_{\textrm{K or $\pi$}}(0)}{m^2 - s + H_{\textrm{K or $\pi$}}(s) - i \sqrt{s} \Gamma(s)},
\end{align}
Here, the subscripts ``GS" and ``KS" refer to the Gounaris-Sakurai \cite{gounaris1968finite} and the K\"{u}hn and Santamaria \cite{kuhn1990tau} parameterizations, respectively.  For the ``GS" version of the BW function, we use the implementation given in Ref.~\cite{kuhn1990tau} with either a Kaon mass cutoff ($BW^{\textrm{GS}}_{\textrm{K}}$) or a pion mass cutoff ($BW^{\textrm{GS}}_{\textrm{$\pi$}}$).  For $BW^{\textrm{KS}}$, the kaon mass cutoff is always used.
As discussed below, the dispersion relation is respected in the GS and constant-width versions of the BW function but is not as closely satisfied in the KS form of the BW function.
\begin{figure*}[!hpt]
	\begin{center}
 	\includegraphics[width=8.6cm]{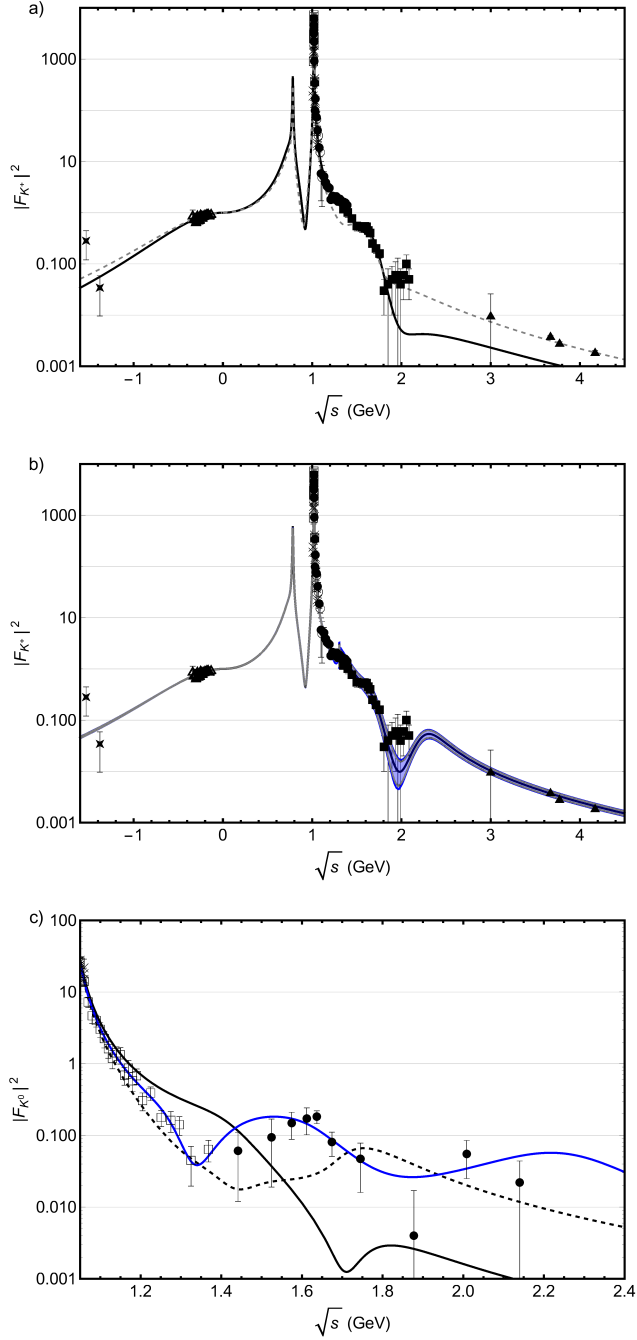}\\
    \caption{\label{fig-fk-fits} Parameterizations of the charged kaon form factor squared along with world data as a function of $\sqrt{s}$ using: (a) parameters from Table 2 of Ref.~\cite{Bruch:2004py} (solid, black) and re-fitted parameters to both spacelike and timelike data using the model of Ref.~\cite{Bruch:2004py} (dashed, black). (b) our best fit (Model 8 in Table~\ref{table-fit-quality}) with the values listed Table~\ref{table-fit-parm} (purple).  The error bands are the 95\% confidence band (gray) and 97.5\% confidence band (blue) taken from model error derived from a multinormal distribution of parameters around their means in Table II. (c) Neutral kaon form factor using the parameterization of Ref.~\cite{Bruch:2004py} (solid, black) along with the re-fitted parameters to both spacelike and timelike data using the model of Ref.~\cite{Bruch:2004py} fit to the new high-$s$ data and spacelike data (dashed, black), and our parameterization from \Eqn{eqn-fk-decomposed} with parameters from Table~\ref{table-fit-parm}.  For (a) and (b), the data are taken from \cite{Achasov:2000am} (crosses), \cite{Akhmetshin:1995vz} (open squares), \cite{Ivanov:1981wf} (full circles), \cite{Dolinsky:1991vq} (open circles), \cite{Bisello:1988ez} (full squares), \cite{Zweber:2006sy} (full triangle $\sim$3 GeV), \cite{Pedlar:2005sj} (full triangle $\sim$3.67 GeV), \cite{Seth:2012nn} (full triangles at $\sim$3.7 and 4.17 GeV), \cite{Amendolia:1986ui} (open triangles), \cite{Coman:2009jk} (full square star).  For (c) the data are from \cite{Akhmetshin:2003rg, Achasov:2000am} (crosses), \cite{Akhmetshin:2002vj} (open squares), \cite{Mane:1980ep} (full circles).}
\end{center}
\end{figure*}

Bruch et al.~\cite{Bruch:2004py} employed the parameterization of Eq.~\ref{eqn-bruch-fk-decomposed} with fixed BW functions for the $\omega$ resonances and KS implementations of the BW functions for $\rho$ and $\phi$ resonances; however, the authors mentioned the possibility of using GS BW functions for the $\rho$ resonances as well. Bruch et al. reported two different fits ("Fit 1" and "Fit 2") to the available data. Fit 1 is constrained by fitting the normalization factors for the $\rho$ resonance only, keeping those of the $\omega$ constant. In Fit 2 the $\rho$ and $\omega$ factors are fitted as independent parameters. These models describe the data up to values of $\sqrt{s}$ $\sim$ 2 GeV$^2$. However, they do not provide a suitable description of the newer, high precision data from Ref.~\cite{Pedlar:2005sj,Seth:2012nn}, under-predicting them by three sigma, as shown in Fig.~\ref{fig-fk-fits}(a). This discrepancy motivated the present analysis and the development of a revised parameterization in order to better represent the available data.

In the first step of the analysis, we attempted to reproduce the models reported in Ref.~\cite{Bruch:2004py} using the same data. The results are shown as Models 1 and 3 in Table~\ref{table-fit-quality} in the third column. We obtain parameter values and fitting statistics in very close agreement with those in Ref.~\cite{Bruch:2004py}. However, as shown in Fig.~\ref{fig-fk-fits}(a), the new high $\sqrt{s}$ data deviate significantly from these models. Thus, the fourth column in Table I shows the results of refitting these models to the extended data set with the new timelike data, while the fifth column shows the results obtained with the entire timelike and spacelike data set. The agreement with the new data is improved by refitting, as shown in Fig.~\ref{fig-fk-fits}(a), but it is clear from Table~\ref{table-fit-quality} that the quality of the fits decreases significantly with the inclusion of the additional data, with $\chi^2$ values increasing by nearly a factor of three. As can be seen in Fig.~\ref{fig-fk-fits}(a), the improvement in the fit at high $\sqrt{s}$ comes at the expense of deviations at lower $\sqrt{s}$. Models 2 and 4 in Table~\ref{table-fit-quality} replace the KS BW functions for the $\rho$ resonances used in Ref.~\cite{Bruch:2004py} by GS BW functions. There is little impact of this for the original data set but substantial improvements in the quality of the fits for the extended data sets. However, the $\chi^2$ values are still about a factor of two larger than was the case for the original fits~\cite{Bruch:2004py}. It is also evident from Table~\ref{table-fit-quality} that most of the increase in $\chi^2$ comes as a result of inclusion of the high $\sqrt{s}$ data rather than inclusion of the spacelike data.

To address these issues and better describe the world kaon form factor data, including
those at higher energies, a new effective parameterization is needed. The simplest method to extend Models 1-4 is to add hadronic resonances. Adding excited $\phi$ and $\omega$ resonances with masses and widths near the Particle Data Group (PDG) values for the higher resonances did not significantly improve the fit. Similarly, adding a "floating'' resonance, with adjustable mass and width, did not improve the fit either. Those models are not reported here. Conversely, adding a higher-order $\rho$ resonance at the PDG values for the mass (2280 MeV$^2$) and width (440 MeV$^2$) did improve the fit; other resonances did not improve the fit as much as the inclusion of this resonance. This is shown in Models 5 and 6, which employ the KS and GS versions, respectively. In both cases, there are significant reductions in $\chi^2$ values over those for Models 1-4. Importantly, the $\chi^2$ value for Model 6 for the entire data set is now close to the values obtained for Models 1 - 4 for the smaller data set. Also of note is that Model 6 with the GS BW function performs significantly better than does Model 5 with the KS BW function.  
Building on Models 5 and 6 and adding another excited $\rho$ resonance with a mass and width as prescribed by the PDG does not provide significant improvement. However, addition of a floating resonance for which we fitted mass and width (Models 7 and 8) gave further reductions in $\chi^2$ values, as shown in Table~\ref{table-fit-quality}. Of these, Model 8, which is the GS BW version, gave the best fit and indeed the best fit of all the models investigated. 

In view of these improvements, the possibility that more resonances might be needed to account for the high-$\sqrt{s}$ data was investigated. However, inclusion of additional resonances
did not further improve the fits. In particular, inclusion of a series of resonances, adapted from the Veneziano amplitude and dual resonance models, suggested in Ref.~\cite{dominguez2001pion} and implemented in Ref.~\cite{Bruch:2004py} in their Eqns. 36, 37, 38, 43, and 44, which resulted in Models 9 and 10, gave worse fits than Models 7 and 8. Additionally, a method of accounting for the high-$\sqrt{s}$ behavior similar to the implementation in Ref.~\cite{Lomon:2016eyp} was also tried but did not provide a better description of the data over the entire $\sqrt{s}$ range as compared to adding broad resonances; those results are not included here. 

In all cases, models with the GS BW functions for the $\rho$ resonances showed improved fits relative to the corresponding models with the KS BW for the $\rho$ resonances (Models 2, 4, 6, 8, and 10 compared with Models 1, 3, 5, 7, and 9, respectively, in Table~\ref{table-fit-quality}).

Finally, a variant of Model 8 in which {\em all} of the BW functions were replaced by the GS versions was investigated as Model 11. As discussed above, such a form has the advantage that it better respects the dispersion relations. However, as shown in Table~\ref{table-fit-quality}, the fit with Model 11 is slightly worse than that for Model 8. 

It is noted that the high-$\sqrt{s}$ data provided the most difficulty for all of these form factor models. Conversely, the spacelike data were generally fit well by all of the models. This is significant because, as discussed in Section~\ref{sec:transverse-density-future}, the higher $\sqrt{s}$ data can have a substantial impact on the form factor.

In summary, of the models investigated, Model 8 provided the best fit to the world kaon form factor data. This model is given by:
\begin{subequations}
\label{eqn-fk-decomposed}
\begin{widetext}
\begin{align}
F_{K^+}(s) &= \frac{1}{2} \Big( c_{\rho} BW^{GS}_{\pi}(m_{\rho},\Gamma_{\rho})(s)+c_{\rho'} BW^{GS}_{\pi}(m_{\rho'},\Gamma_{\rho'})(s)+c_{\rho''} BW^{GS}_{\pi}(m_{\rho''},\Gamma_{\rho''})(s) \nonumber\\ 
&+ c_{\rho'''} BW^{GS}_{\pi}(m_{\rho'''},\Gamma_{\rho'''})(s) + c_{\rho^{''''}} BW^{GS}_{\pi}(m_{\rho^{''''}},\Gamma_{\rho^{''''}})(s) \Big) +\nonumber\\
&+\frac{1}{6} \Big( c_{\omega} BW(m_{\omega},\Gamma_{\omega})(s)+c_{\omega'} BW(m_{\omega'},\Gamma_{\omega'})(s)+c_{\omega''} BW(m_{\omega''},\Gamma_{\omega''})(s) \big) \nonumber\\
&+\frac{1}{3} \Big( c_{\phi} BW^{KS}(m_{\phi},\Gamma_{\phi})(s)+c_{\phi'} BW^{KS}(m_{\phi'},\Gamma_{\phi'})(s) \Big)\\
F_{K^0}(s) &= \frac{-1}{2} \Big( c_{\rho} BW^{GS}_{\pi}(m_{\rho},\Gamma_{\rho})(s)+c_{\rho'} BW^{GS}_{\pi}(m_{\rho'},\Gamma_{\rho'})(s)+c_{\rho''} BW^{GS}_{\pi}(m_{\rho''},\Gamma_{\rho''})(s) \nonumber\\ 
&+ c_{\rho'''} BW^{GS}_{\pi}(m_{\rho'''},\Gamma_{\rho'''})(s) + c_{\rho^{''''}} BW^{GS}_{\pi}(m_{\rho^{''''}},\Gamma_{\rho^{''''}})(s) \Big) +\nonumber\\
&+\frac{1}{6} \Big( c_{\omega} BW(m_{\omega},\Gamma_{\omega})(s)+c_{\omega'} BW(m_{\omega'},\Gamma_{\omega'})(s)+c_{\omega''} BW(m_{\omega''},\Gamma_{\omega''})(s) \big) \nonumber\\
&+\frac{1}{3} \Big(\eta_\phi c_{\phi} BW^{KS}(m_{\phi},\Gamma_{\phi})(s)+c_{\phi'} BW^{KS}(m_{\phi'},\Gamma_{\phi'})(s) \Big)
\end{align}
\end{widetext}
\end{subequations}
The coefficients $c$ are as in \Eqn{eqn-bruch-fk-decomposed}.  The Breit-Wigner functions are defined in \Eqns{BWfixed}, (\ref{BWKS}), and (\ref{BWGS}). The coefficient $\eta_\phi$ is a fixed constant listed in Table~\ref{table-fit-parm}. 

Both the width and position of the peak of the ground state $\phi$ resonance along with the 4th $\rho$ resonance were optimized using $\chi^2$ minimization.  All other widths and positions of the peaks were PDG values.

To include neutral kaon and charged kaon data, a flag parameter, $f$, (either 1 for charged kaon data or 0 neutral kaon data) was introduced to augment the data values, and all of the spacelike and timelike data for both the charged and neutral kaon form factors were used to fit to an expression $f F_{K^+}(s) + (1-f) F_{K^0}(s)$, so all data could be fit to the same parameters.

We have fitted the parameters of \Eqn{eqn-fk-decomposed} to the existing and new timelike kaon form factor data from Ref.~\cite{Pedlar:2005sj,Seth:2012nn} as well as existing spacelike data. The resulting fit parameters are listed in Table~\ref{table-fit-parm} and Figs.~\ref{fig-fk-fits}(b), (c) illustrate our parameterization of the charged and neutral kaon form factor along with the one from Ref.~\cite{Bruch:2004py}. To evaluate the effect of the fitted parameters on the form factor, the parameters were varied following a Gaussian distribution around their central values while all other non-fit parameters were held fixed. The resulting distribution of form factor values for a fixed $Q^2$ provided a distribution where a 95\% and 97.5\% confidence bands may be computed.  \Figref{fig-fk-fits} (b) contains both bands for the best fit model. The parameters for the best fit are listed in Table~\ref{table-fit-parm}. 
\begin{figure}[!hpt]
	\begin{center}
 	\includegraphics[width=8.6cm]{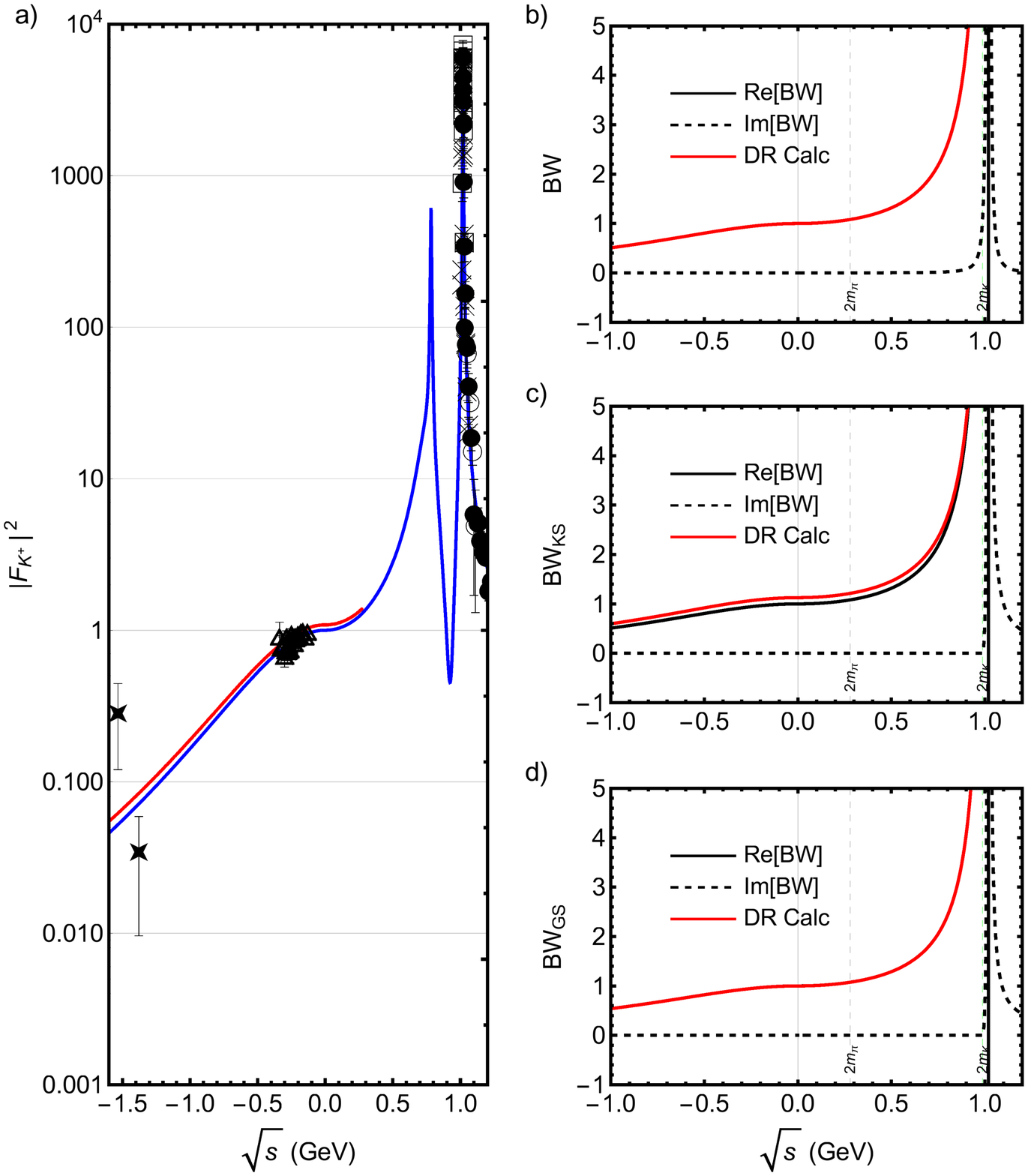}\\
    \caption{\label{fig:continuation} (a) Comparison of the spacelike dispersion relation of \Eqn{eqn:dr} (red curve) and the analytic continuation of our parameterization from \Eqn{eqn-fk-decomposed} into the spacelike region (blue curve). The data are the same as in \Figref{fig-fk-fits}. The panels on the right side show a dispersion relation calculation (red curve), the real parts of the Breit-Wigner (BW) functions (black curve), and the imaginary parts of the BW functions (black dashed curve) for: (b) A single constant-width BW with its dispersion relation calculation (mass 1020 MeV and width 4.36 MeV), (b) A single KS BW with an $s$-dependent width and a cutoff of $2 m_{K^+}$ (mass 1020 MeV and width 4.36 MeV), and (d) A single GS BW function with an $s$-dependent width and a cutoff of $2 m_{K^+}$ (mass 1020 MeV and width 4.36 MeV).}
\end{center}
\end{figure}
\begin{small}
  \begin{table*}
\caption{\it Comparison of $\chi^2$ and number of degrees of freedom (DOF) for various parameterizations of the kaon form factor. Our parameterization used for the extraction of the transverse density is Model 8.  All models (except Model 11) use fixed BW functions for the $\omega$ resonances, and KS implementations of BW functions for $\phi$ resonances.  In the description of the model, ``GS" or ``KS" refers to the BW function for the $\rho$ resonances.  For Model 11, \emph{all} BW functions for all resonances are the GS implementation.}
\label{table-fit-quality} \smallskip
\small
%\begin{ruledtabular}
\begin{tabular}{||l|l|c|c|c||} 
\hline
\multicolumn{1}{||l|}{\multirow{2}{*}{Model Number}} & \multicolumn{1}{|l|}{\multirow{2}{*}{Description}} &  Data available        & All Timelike   & All Data    \\
                                             & & to Bruch \emph{et al.} & Data           &(including Spacelike data) \\
                                         &     &  $\chi^2/$DOF          & $\chi^2/$DOF   & $\chi^2/$DOF   \\
\hline
 1 & Bruch KS,& \multicolumn{1}{c|}{\multirow{2}{*}{346/242}} & \multicolumn{1}{c|}{\multirow{2}{*}{881/246}} & \multicolumn{1}{c||}{\multirow{2}{*}{904/273}}  \\
 & \phantom{$\,\,\,\,$}Fit 1 & & & \\ 
 2 &  Bruch GS,& \multicolumn{1}{c|}{\multirow{2}{*}{365/242}} & \multicolumn{1}{c|}{\multirow{2}{*}{616/246}} & \multicolumn{1}{c||}{\multirow{2}{*}{640/273}}  \\
& \phantom{$\,\,\,\,$}Fit 1 & & & \\ 
 3 &  Bruch KS,& \multicolumn{1}{c|}{\multirow{2}{*}{292/240}} & \multicolumn{1}{c|}{\multirow{2}{*}{852/244}} & \multicolumn{1}{c||}{\multirow{2}{*}{876/271}}  \\
& \phantom{$\,\,\,\,$}Fit 2 & & & \\ 
 4 &  Bruch GS,& \multicolumn{1}{c|}{\multirow{2}{*}{288/240}} & \multicolumn{1}{c|}{\multirow{2}{*}{614/244}} & \multicolumn{1}{c||}{\multirow{2}{*}{638/271}}  \\
& \phantom{$\,\,\,\,$}Fit 2 & & & \\ 
 5 &  Bruch KS w/ & & \multicolumn{1}{c|}{\multirow{2}{*}{482/243}} & \multicolumn{1}{c||}{\multirow{2}{*}{505/270}}  \\
&\phantom{$\,\,\,\,$}Added $\rho$(2280,440) & & & \\ 
 6 &  Bruch GS w/ & & \multicolumn{1}{c|}{\multirow{2}{*}{322/240}} & \multicolumn{1}{c||}{\multirow{2}{*}{346/270}}  \\
&\phantom{$\,\,\,\,$}Added $\rho$(2280,440) & & & \\ 
 7 &  Bruch KS w/ 2 Added $\rho$s& & \multicolumn{1}{c|}{\multirow{2}{*}{284/240}} & \multicolumn{1}{c||}{\multirow{2}{*}{307/267}}  \\
&\phantom{$\,\,\,\,$}($\rho$(2280,440) and Varied)& & & \\ 
\hline
 8 Best &  Bruch GS w/ 2 Added $\rho$s & & \multicolumn{1}{c|}{\multirow{2}{*}{267/240}} & \multicolumn{1}{c||}{\multirow{2}{*}{290/267}}  \\
&\phantom{$\,\,\,\,$}($\rho$(2280,440) and Varied)& & & \\ 
\hline
 9 &  Bruch KS w/ & & \multicolumn{1}{c|}{\multirow{2}{*}{482/243}} & \multicolumn{1}{c||}{\multirow{2}{*}{506/270}}  \\
& \phantom{$\,\,\,\,$}Added $\rho$ series & & & \\ 
 10 &  Bruch GS w/ & & \multicolumn{1}{c|}{\multirow{2}{*}{446/243}} & \multicolumn{1}{c||}{\multirow{2}{*}{470/270}}  \\
& \phantom{$\,\,\,\,$}Added $\rho$ series & & & \\
 11 &  All GS w/ & & \multicolumn{1}{c|}{\multirow{2}{*}{278/240}} & \multicolumn{1}{c||}{\multirow{2}{*}{302/267}}  \\
& \phantom{$\,\,\,\,$}2 Added $\rho$s & & & \\ 
\hline
\end{tabular}
%\end{ruledtabular}
 \end{table*}
\end{small}
\begin{small}
\begin{table*}
\caption{\it Fit parameters and uncertainties of our best fit (Model 8 in Table~\ref{table-fit-quality}) for both charged and neutral kaon form factor. }
\label{table-fit-parm} \smallskip
\small
\begin{tabular}{||l|c|c|c||} 
\hline
 Model       &    \multicolumn{1}{c|}{\multirow{2}{*}{Input}}   & \multicolumn{1}{c|}{\multirow{2}{*}{Estimate}}     & Standard   \\
  Parameter  &                                                    &                                                  &  Error     \\
             & MeV                                                & MeV                                              & MeV        \\
\hline
 $m_{\phi}$                                     &       -         &      1019.3      &         0.02       \\
 $\Gamma_{\phi}$                                &       -         &      4.23        &         0.04       \\
 $m_{\phi^\prime}$                              &   1680          &      -           &         -          \\
 $\Gamma_{\phi^\prime}$                         &   150           &      -           &         -          \\
\hline
 $m_{\rho}$                                     &   775           &      -           &         -          \\
 $\Gamma_{\rho}$                                &   150           &      -           &         -          \\
 $m_{\rho'}$                                    &   1465          &      -           &         -          \\
 $\Gamma_{\rho'}$                               &   400           &      -           &         -          \\
 $m_{\rho''}$                                   &   1720          &      -           &         -          \\
 $\Gamma_{\rho''}$                              &   250           &      -           &         -          \\
 $m_{\rho'''}$                                  &   2280          &      -           &         -          \\
 $\Gamma_{\rho'''}$                             &   440           &      -           &         -          \\
 $m_{\rho^{\emph{\textrm{iv}}}}$                &       -         &      1294        &         16         \\
 $\Gamma_{\rho^{\emph{\textrm{iv}}}}$           &       -         &      174         &         60         \\
\hline
 $m_{\omega}$                                   &   783           &      -           &         -          \\
 $\Gamma_{\omega}$                              &   8.4           &      -           &         -          \\
 $m_{\omega'}$                                  &   1425          &      -           &         -          \\
 $\Gamma_{\omega'}$                             &   215           &      -           &         -          \\
 $m_{\omega''}$                                 &   1670          &      -           &         -          \\
 $\Gamma_{\omega''}$                            &   315           &      -           &         -          \\
\hline
\hline
 $c_{\phi}$                                     &       -         &      0.99        &         0.01       \\
 $c_{\phi'}$                                    &   $1-c_{\phi}$  &      -           &         -          \\
\hline
 $c_{\rho}$                                     &       -         &      1.06        &         0.01       \\
 $c_{\rho'}$                                    &       -         &      -0.18       &         0.02       \\
 $c_{\rho''}$                                   &       -         &      -0.02       &         0.006      \\
 $c_{\rho'''}$                                  &       -         &      0.08        &         0.004      \\
 $c_{\rho^{\emph{\textrm{iv}}}}$                &   $1-(c_{\rho}+c_{\rho'}+c_{\rho''}+c_{\rho'''})$  &      -           &         -          \\
\hline
 $c_{\omega}$                                   &       -         &      1.06        &         0.01       \\
 $c_{\omega'}$                                  &       -         &      -0.18       &         0.02       \\
 $c_{\omega''}$                                 &   $1-(c_{\omega}+c_{\omega'})$         &      -0.02       &         0.006      \\
\hline
 $n_{\phi}$                                     &   1.011         &      -           &         -          \\
\hline
$\chi^2$/dof &  \multicolumn{3}{c||}{\multirow{1}{*}{290/267}} \\
\hline
\end{tabular}
\end{table*}
\end{small}

For all of the models we considered we continued the form factor parameterization into the spacelike region $s <$ 0, as described below, and included the available spacelike data in the fits.
In general, the analytic continuation can be carried out using dispersion relations based on the Kramer-Kronig relations. Using only the imaginary part of a generic function guarantees regularized analytic continuation. Here, the Breit-Wigner formulas in the dispersion representation are of the form
\begin{equation}\label{eqn:dr}
 \textrm{Re} \left\{f(t)\right\}=\frac{1}{\pi} \mathscr{P} \int_{-\infty}^{\infty} \frac{\textrm{Im} \left\{f(s)\right\}}{s-t} ds,
\end{equation}
%
%${\mathop{\rm Im}\nolimits}$
where \textrm{Re} and \textrm{Im} are the real and imaginary parts, and $\mathscr{P}$ is the Cauchy Principal Value. Since the function is evaluated at a spacelike point ($t<0$) and the form factor is real on the spacelike domain, we can restrict the integral to the 
$s>0$ region.

In our procedure, we evaluate the Breit-Wigner functions of the form factor in the negative real argument using the simplest branch. To check the causality of our analytically continued Breit-Wigner function, we compared our results to those obtained with the dispersion relation. The results are shown in \Figref{fig:continuation}. The constant-width Breit-Wigner and GS parameterizations are in agreement with the dispersion relation. The KS parameterization deviates on the 10\% level. Overall, the analytically continued parameterization provides a good description of both timelike and spacelike kaon form factor data. The 10\% deviation on the spacelike side is due to the KS parameterization used for the $\phi$ resonance in \Eqn{eqn-fk-decomposed}. We further evaluate the impact of the spacelike data on the transverse density in section~\ref{sec:uncertainty-experimental}.

\section{Extraction of the transverse charge density}
\label{sec:rho_extr}

The kaon transverse charge density $\rho_{K}(b)$ is defined as the two-dimensional Fourier transform of the spacelike kaon form factor,
\begin{equation}
\label{eqn-charge-density}
\rho_{K} (b) = \int_0^{\infty} \frac{\mathrm{d}Q}{2 \pi} Q J_0(Qb) F_{K} (t=-Q^2),
\end{equation}
where $Q$ is the square root of the four-momentum transfer, $J_0$ is the Bessel function, and $F_{K}$ is a function of the Mandelstam variable $-t$. The function $\rho_K$ describes the probability that a charge is located at a transverse distance $b$ from the transverse center of momentum in the nucleon. It is normalized as $\int \mathrm{d}^2 b \rho_{K}(b)=1$. Equation~\ref{eqn-charge-density} can be used to extract the transverse charge density from spacelike kaon form factor data, as was done for the pion in Ref.~\cite{Carmignotto:2014rqa}. However, spacelike kaon form factor data are very sparse, and we thus extract the transverse density from timelike kaon form factor data using a dispersion representation. 

The singularities of $F_K(s)$, which is an analytic function of $s$, are confined to a cut along the positive real axis starting at the threshold value $s=4m_K^2$. With this the kaon form factor of the kaon can be written,
\begin{equation}
\label{eqn-form-factor}
F_K(t) = \int_{4m_K^2}^{\infty} \frac{\mathrm{d} t^{\prime}}{t^\prime-t-i\epsilon} \frac{\textrm{Im}(F_K(t^{\prime}))}{\pi}.
\end{equation}
Perturbative QCD predicts that $F_K(t) \sim \alpha_s(t)/|t|$ as $t \rightarrow \infty$. This allows one to use an unsubtracted dispersion relation as described in Ref.~\cite{Lomon:2016eyp}. Substituting \Eqn{eqn-form-factor} into (\ref{eqn-charge-density}) one obtains
\begin{equation}
\label{eqn-kaon-trans-density}
\rho_K(b) = \int_{4m_K^2}^{\infty} \frac{\mathrm{d} t}{2 \pi} K_0(\sqrt{t} b) \frac{\textrm{Im}(F_K(t+i0))}{\pi},
\end{equation}
where $\textrm{Im}(F_K(t))$ is the imaginary part (spectral function) of the kaon form factor weighted by $K_0$, the modified Bessel function. At large values of $t$, $K_0$ decreases exponentially, so that the spectral function samples only values $\sqrt{t} \sim 1/b$ at a given transverse distance $b$. 

The physical region for the kaon timelike form factor starts at $t=4 m_K^2$, and thus experimental data are available for the region above $t \sim$ 1 GeV$^2$. High-quality $e^+ e^-$ annihiliation data exist for values of $\sqrt{t}$ up to 2 GeV and new data have become available up to 4.2 GeV. This allows for determination of the kaon transverse charge density to values of $b$ down to $b \sim$ 0.05 fm. 

The extraction of the kaon transverse density requires as input the experimental value of $F_{K}$ obtained from the parameterization shown in \Figref{fig-fk-fits}(b). The uncertainty on the extraction thus also depends on the experimental uncertainties. The total uncertainty on $\rho_{K}(b)$ has two main sources: 1) experimental uncertainties on the individual measurements and combining data from different experiments in the region where data exist and 2) uncertainties due to the lack of data in the region beyond $\sqrt{t}>$4.2 GeV, where no measurements exist. The experimental uncertainties are taken into account directly in the coefficients $c_n$ through \Eqn{eqn-fk-decomposed}. However, uncertainty due to lack of timelike kaon form factor data for values of $Q^2>Q^2_{max}$=17.4 GeV$^2$ must also be estimated. Both sources of uncertainty are discussed next.

\subsection{Experimental Uncertainty}
\label{sec:uncertainty-experimental}

The dispersion integral in \Eqn{eqn-kaon-trans-density} includes a parameterization of the kaon form factor data and a weight factor. Uncertainties from the kaon form factor data were used to estimate the uncertainty in $\rho_K$. In particular, the uncertainty on $F_K$ directly results in an uncertainty on the coefficients $c_n$ and thus directly contributes to $\rho_{K}(b)$. 

\begin{figure}[!hpt]
	\begin{center}
 	\includegraphics[width=8.6cm]{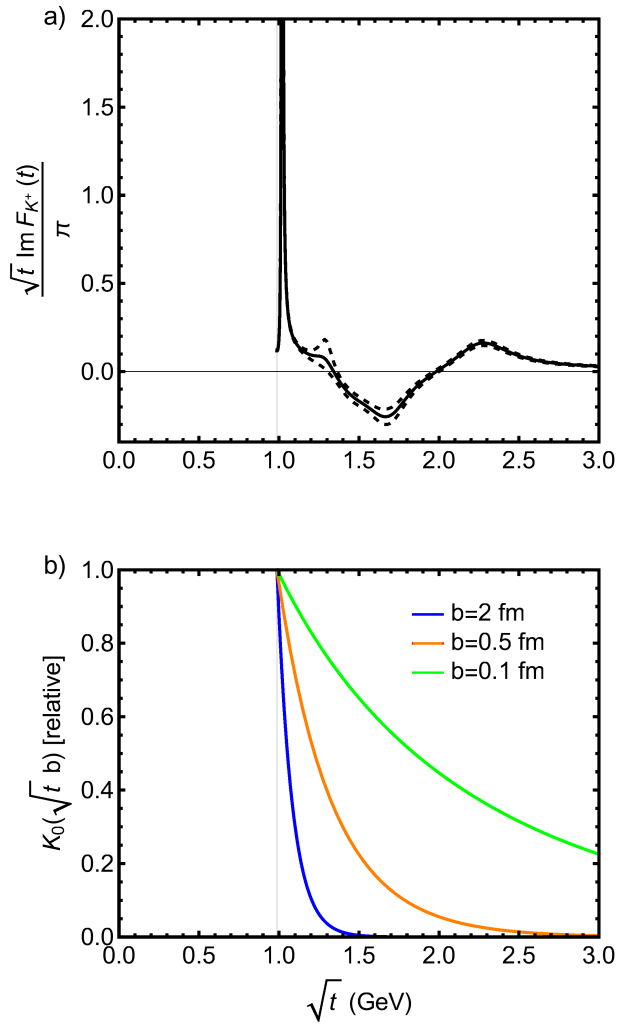}\\
    \caption{\label{fig:fk-imaginary-weight} (a) The imaginary part of the kaon form factor obtained from our best fit whose parameters are listed in Table~\ref{table-fit-parm} as a function of $\sqrt{t}$ (solid black curve). The threshold energy is $\sqrt{t}$=$2m_K$ and indicated by the gray vertical line. One sigma confidence bands are included as dashed curves. (b) The weight factor $K_0(\sqrt{t} b)$ in the dispersion representation of the transverse charge density in \Eqn{eqn-kaon-trans-density} as a function of $\sqrt{t}$ for a several values of $b$=0.1 fm (green), 0.5 fm (orange), 2 fm (blue).}
\end{center}
\end{figure}

The imaginary part of the form factor calculated using our best fit (Model 8 of Table~\ref{table-fit-quality}) is shown in \Figref{fig:fk-imaginary-weight}a. The dominance of the $\phi$ pole in $\textrm{Im}(F_K)$ and the alternating sign of successive resonance contributions at larger values of $\sqrt{t}$, as expected from theoretical considerations, can be seen as well. We estimated the statistical uncertainty assuming uncorrelated uncertainties in the fit parameters. The dashed lines in \Figref{fig:fk-imaginary-weight}a show the resulting 1$\sigma$ error band.  

The variance in the $\phi$ meson mass region is at the few percent level. At energies above 1 GeV it becomes larger, reaching the size of its value at $\sqrt{t}$=1.3 GeV. However, in this energy region, an uncorrelated estimate is likely an upper bound of the uncertainty. Correlations between statistical fluctuations of the coupling and width of higher resonances would reduce the overall fluctuations of the imaginary part. For energies above 4.2 GeV one cannot reliably estimate the relative uncertainty of the imaginary part using this method as no data are available to constrain the fit.  However, the imaginary part is very small in this region and contributes little to the transverse charge density at $b>$0.1 fm as discussed below.

To understand the relative importance of the uncertainty on the imaginary part to the charge density we evaluated the weight factor of the dispersion integral as a function of $\sqrt{s}$. Figure~\ref{fig:fk-imaginary-weight}b shows the weight factor for several values of $b$ normalized to the same value at threshold, $\sqrt{s}$=$2m_K$. The effective distribution of strength in $\sqrt{s}$ has a strong dependence on $b$. For example, at $b$=0.1 fm a substantial contribution to the dispersion integral comes from the region $\sqrt{s}>$ 1 GeV, where the parameterization of $\textrm{Im}(F_\pi(s))$ shows considerable uncertainty. At $b$=0.5 fm these contributions are reduced and effectively suppressed for $\sqrt{t}>$ 3 GeV. This implies perfect vector meson dominance in the dispersion integral. At large distances of $b \sim$ 2 fm, one begins to suppress the $\phi$ mass region and emphasizes the near-threshold region of the form factor, $\sqrt{t}$=2$m_K$.

\begin{figure}[!hpt]
	\begin{center}
 	\includegraphics[width=8.6cm]{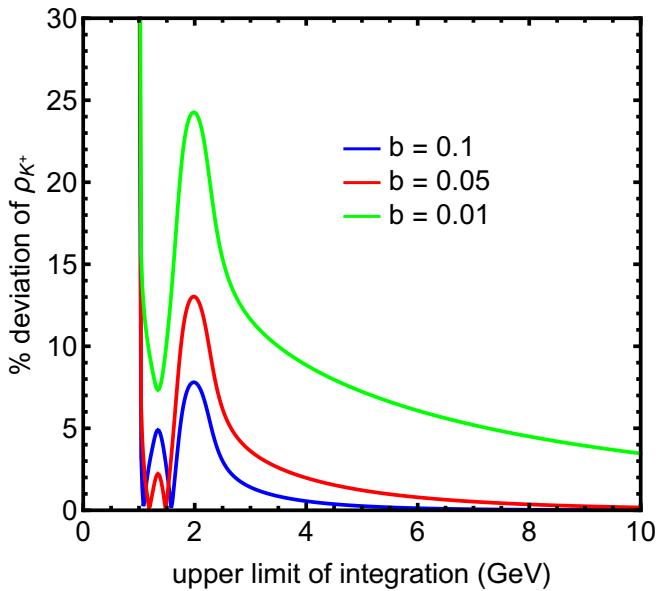}\\
    \caption{\label{fig:fk_trans_density_convergence} Percentage deviation of the dispersion integral of \Eqn{eqn-kaon-trans-density} as a function of the upper limit integration cutoff. Shown is the contribution to the integral for different transverse distances, $b$=0.01, 0.05, and 0.1 fm. The integrand is evaluated using the parameterization shown in \Eqn{eqn-fk-decomposed} with parameters from Table~\ref{table-fit-parm}.}
\end{center}
\end{figure}

To quantify uncertainties for values of $Q^2>$17.4 GeV$^2$ ($\sqrt{t}$=4.2 GeV) where no measurements exist, we studied the numerical convergence of the $\sqrt{s}$ dispersion integral for different upper limits. Figure \ref{fig:fk_trans_density_convergence} shows the percentage deviation of the transverse density from its value when integrated to infinity for different cutoffs applied to the upper limit of the $\sqrt{t}$ integral of \Eqn{eqn-kaon-trans-density}. Here, the integral is evaluated with our best fit and its parameter values from Table~\ref{table-fit-parm}. At $b$=0.1 fm and assuming a 100\% uncertainty, the region $\sqrt{s}>$ 4 GeV accounts for $<$ 1\% of the total integral. A change of the spectral function in this region by a factor of 2-3 from its nominal value would change the transverse density by at most 2-3\%. The error in the transverse density is thus dominated by the mass region of 1$<\sqrt{t}<$4 GeV for which we have estimated the experimenental uncertainty in section~\ref{sec:uncertainty-experimental}. With a 100\% uncertainty at $\sqrt{t}$=2 GeV, where the integral has converged within 8\% of its value, one would expect an uncertainty of the density of at least 8\%. At smaller distances to about $b$=0.05 fm, the region $\sqrt{t}>$ 4 still contributes very little. While the integral requires larger values of $\sqrt{t}$ to converge, the contribution from $\sqrt{t}>$ 4 GeV is still only $\sim$ 2\% and the overall uncertainty is dominated by the region 1$<\sqrt{t}<$4 GeV. At distances of $b$=0.01 fm, the contribution of the region $\sqrt{t}>$ 4 GeV increases to 10\%. At $\sqrt{t}$=2 GeV, where the integral has converged to about 25\% of its value, and thus one would expect an uncertainty of the density of at least 25\%. At larger values of $\sqrt{t}$, e.g. at $b$=0.5 fm, the integral has already converged at $\sqrt{t}$=1 GeV and the overall uncertainty is dominated by the low energy region. In this region the parameter errors are so small that the model dependence of the parameterizations cannot be neglected any longer. 

The model dependence was studied by extracting the transverse density for form factor parameterizations that describe the data equally well overall, but have different characteristics in the regions $\sqrt{t}<$ 1 GeV, 1 GeV$< \sqrt{t}<$ 4 GeV, and $\sqrt{t}>$ 4 GeV. We also compared the impact of adding resonances of different masses and widths and a perturbative form factor behavior to our nominal parameterization. The effect of adding a resonance at $\sqrt{s}$=4 GeV and width 0.001 GeV compared to a resonance of width 1 GeV at the same center of mass energy is $\sim$ 10\% at $b$=0.05 fm on the transverse density. The individual differences in the extracted density compared to our nominal parameterization are 17\% (narrow resonance) and 5\% (broad resonance), respectively.

\subsection{Transverse Charge Density}
\label{sec:transverse-density}

\begin{figure}[!hpt]
	\begin{center}
% 	\includegraphics[width=8.6cm]{./fig5}\\
 	\includegraphics[width=8.6cm]{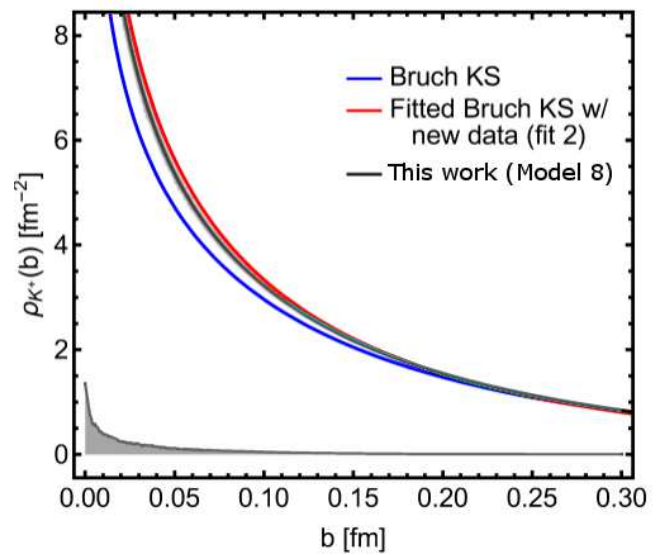}\\
    \caption{\label{fig:fk_trans_density_modeldep} Transverse charge density of the charged kaon. The solid line indicates the dispersion integral of \Eqn{eqn-kaon-trans-density} evaluated with the form factor parameterization and fit parameters from Ref.~\cite{Bruch:2004py} (blue curve), with re-fitted parameters using the parameterization of Ref.~\cite{Bruch:2004py}(red curve), and our parameterization of \Eqn{eqn-fk-decomposed} with parameters from Table~\ref{table-fit-parm} (black curve). The gray band indicates a 95\% confidence band of the reparameterization with parameters selected from a multinormal distribution.  The shaded gray curve near the origin is a plot of the 95\% confidence band width.}
\end{center}
\end{figure}

We turn now to our stated  goal of using the world data on the timelike kaon form factor to extract the kaon transverse charge density. Fig.~\ref{fig:fk_trans_density_modeldep} shows the result obtained from the dispersion integral and our parameterization of the kaon form factor. The transverse density rises rapidly at small values of $b$ and shows an exponential fall off at larger distances. This behavior appears to be consistent with a central density having a logarithmic divergence as $b$ approaches the origin. However, the divergence in the density may be a result of using a simple parameterization not well constrained at small values of $b$ (large values of $Q^2$). As an illustration of the impact of constraining the parameterization  we show the transverse densities calculated from (i) the model of Ref.~\cite{Bruch:2004py} (data to $\sqrt{t} \sim$ 2 GeV), (ii) the model of Ref.~\cite{Bruch:2004py} refitted with new data (to $\sqrt{t}$=4.2 GeV), and (iii) Model 8 from the present work. One can see that the impact is $b$ dependent, increasing from about 14\% at $b$=0.1 fm to 25\% at $b$=0.05 fm. We discuss the impact of future planned data in section~\ref{sec:transverse-density-future}.

\subsection{Nucleon Meson Cloud and Kaon Charge Density}
\label{sec:transverse-density-compare}

\begin{figure}[!hpt]
	\begin{center}
 	\includegraphics[width=8.6cm]{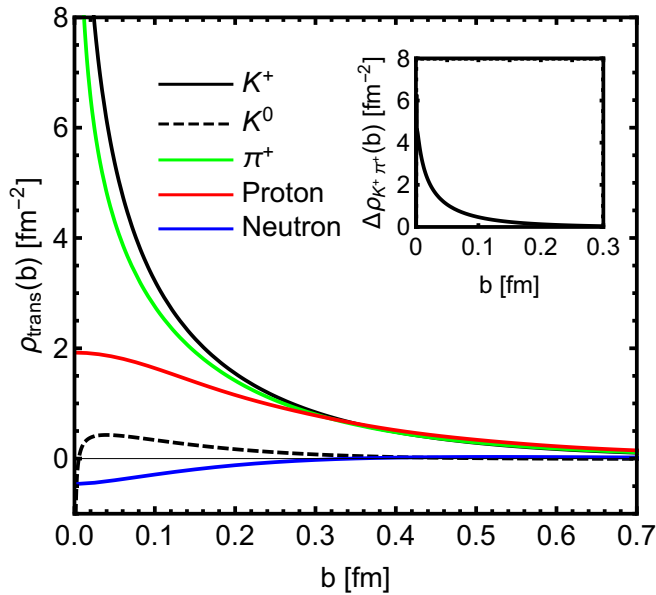}\\
    \caption{\label{fig:rhok_spacelike_current} The transverse charge density of the charged and neutral kaon (black curve and black dashed curve, respectively) calculated from \Eqn{eqn-kaon-trans-density} compared with the charged pion (green curve), the proton (red curve), and the neutron (blue curve).  The inset shows the difference of the charged kaon and the charged pion.}
\end{center}
\end{figure}

A recent work~\cite{Strikman:2010pu} explored the proton transverse charge density finding that the non-chiral core is dominant up to relatively large distances of $\sim$~2~fm. This suggests that there is a non-pionic core of the proton, as one would obtain in the constituent quark or vector meson dominance models. One does not usually think of the kaon or pion having a meson cloud since a, {\it e.g.}, $\rho\pi$ component would involve a high excitation energy. Therefore it is interesting to compare the proton, kaon, and pion transverse charge densities. This is done in Fig.~\ref{fig:rhok_spacelike_current}.

For values of $b$ less than about 0.3~fm the transverse charge density  of the charged kaon is larger than that of the pion and proton. This higher density might be expected because the kaon's radius of $0.34$~fm is smaller than that of the pion ($0.672$~fm) and the proton ($0.84$~fm). As previously noted~\cite{Miller:2009qu,Miller:2011du}, it is possible that both the pion and kaon's transverse density is singular for small values of $b$. An interesting feature is that the curves seem to coalesce in the region $b>$~0.3~fm (at least within current uncertainties). 

The neutral kaon density peaks around $b$=0.02 fm and then rapidly drops to negative values as $b$ approaches the origin. It is about the same as that of the neutron, which peaks at about $b$=0.5~fm and more slowly approaches negative values at the center. As discussed in Refs.~\cite{Miller:2007uy,Miller:2008jc,Rinehimer:2009sz}, if the neutron were sometimes a proton surrounded by a negatively charged pionic cloud, the central charge density should be positive. However, a negative charge density can be explained by the dominance of the neutron's $d$ quarks at high values of $x$ leading to a negative contribution to the charge density, which becomes localized near the center of mass of the neutron. The $d$ quark in the neutral kaon may have a similar impact. The curves come together for values $b>$~0.3~fm (within current uncertainties).

\section{Impact of future experiments}
\label{sec:transverse-density-future}

\begin{figure}[!hpt]
	\begin{center}
 	\includegraphics[width=8.6cm]{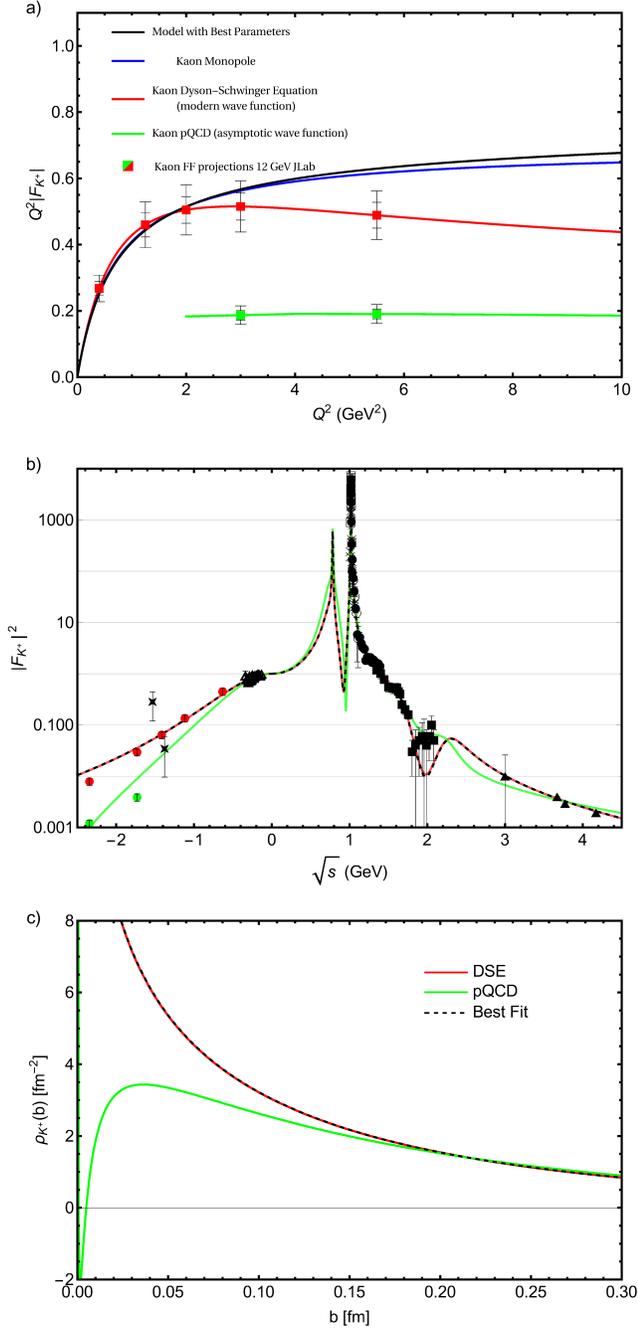}\\
    \caption{\label{fig:trans_density_compare} Investigation of the impact of new spacelike data projected for the 12 GeV JLab along with their uncertainties.  (a) The values of the projected points calculated in the framework of the Dyson-Schwinger Equation (DSE) (red points and curve), and perturbative QCD (pQCD) (green curve).  Also shown are the monopole curve (blue) and our parameterization with values from Table~\ref{table-fit-parm} (black curve).  (b) The impact of the projected data points on our parameterization of the kaon form factor. The red curve uses the values of the form factor calculated in the DSE framework and the green curve those calculated in the pQCD framework (green points).  The black dashed curve is \Eqn{eqn-fk-decomposed} with parameters from Table~\ref{table-fit-parm}. (c) Impact of the projected points on the transverse charge density of \Eqn{eqn-kaon-trans-density}. The difference gives an estimate of the model uncertainty.}
\end{center}
\end{figure}

The extraction of the kaon transverse density from timelike form factor data is complicated by the fact that the relative strength of the continuum is largely unknown. Measurements of the kaon form factor to higher $Q^2$ in the spacelike regime may shed light on this aspect. Experiments at the 12 GeV JLab~\cite{E12-07-105,E12-09-011} have the potential to extend the $Q^2$ range of spacelike kaon form factor data to $Q^2$ $\sim$5.5 GeV$^2$ as illustrated in Fig.~\ref{fig:trans_density_compare}(a). This should be large enough to resolve differences between calculation and the monopole fit, or rule out both. The envisioned Electron-Ion Collider (EIC) has the potential to further extend this reach to about 23 GeV$^2$.

Assuming that all data from 12~GeV JLab are measured, we analyze the possible impact of the new data on the precision of the extraction of the kaon charge distribution. The results are shown in Fig.~\ref{fig:trans_density_compare}(b) and (c). If the new form factor data were described by the calculation in the Dyson-Schwinger (DSE) framework~\cite{Gao:2017mmp,Chen:2016sno,Burden:1995ve} or the monopole fit, the transverse density would follow that obtained with our parameterization. 
%The overall impact on the value of the transverse density is on the order of {\bf TH: QUANTIFY} XYZ\% at small $b$. 
If a perturbative QCD model, e.g., that of Ref.~\cite{Bakulev:2004cu} with asymptotic wave function, described the data, the transverse density would approach the origin slowly, peak at about $b$=0.02~fm, and diverge rapidly towards negative values. The difference in the transverse density obtained with these two models gives an estimate of the size of the uncertainty, albeit very conservative, in the transverse density as $b$ approaches zero. 

\section{Summary}
\label{sec:sum}

In this paper we used the world data on the timelike kaon form factor to extract the transverse kaon charge density. Recent measurements from CLEO extended timelike kaon form factor data into the region $\sqrt{s}$=3-4 GeV and thus allow access to the region of short transverse distances. We created a superset of timelike and spacelike kaon form factor data and developed a parameterization that describe it. For the spacelike kaon form factor data we extracted two new data points to further constrain our parameterization. With the kinematic reach of the available form factor data and the uncertainties in separating real and imaginary parts we estimate the uncertainty on the resulting transverse density to be 5\% at $b$=0.025~fm and significantly better at larger distances. New kaon data planned with the 12 GeV Jefferson Lab may have a significant impact on the charge density at distances of $b<$0.1 fm.

%%% ACKNOWLEDGMENTS %%%
\vspace{0.6in}
\centerline{ACKNOWLEDGMENTS} 

We are grateful for constructive and instructive remarks from Craig Roberts and Ian Cloet. This work was supported in part by NSF grants PHY-1306227 and PHY-1306418, and USDOE Grant no. DE-FG02-97ER-41014.

%%% BIBLIOGRAPHY %%%

\bibliography{thbibtanja}

\end{document}